\documentclass[9pt,twocolumn,twoside]{optica}
\setboolean{shortarticle}{false}
\setboolean{minireview}{false}
\usepackage{mathtools}
\usepackage[font=small,labelfont=bf,justification=justified]{caption}
\usepackage{pdfpages}

\title{Acousto-optic tomography beyond the acoustic diffraction-limit using speckle decorrelation}

\author[1]{Daniel Doktofsky}
\author[1]{Moriya Rosenfeld}
\author[1,*]{Ori Katz}

\affil[1]{Applied Physics Department, Hebrew University of Jerusalem, 9190401 Jerusalem, Israel.}

\affil[*]{Corresponding author: orik@mail.huji.ac.il}

\begin{abstract}
Acousto-optic tomography (AOT) enables optical-contrast imaging deep inside scattering samples via localized ultrasound modulation of scattered light. However, the resolution of AOT is inherently limited by the ultrasound focus size, prohibiting microscopic investigations. In the last few years, advances in the field of digital wavefront-shaping have allowed the development of novel approaches for overcoming the acoustic resolution limit of AOT. However, these novel approaches require the execution of thousands of wavefront measurements within the sample speckle decorrelation time, limiting their application to static samples.
Here, we show that it is possible to surpass the acoustic resolution limit with a conventional AOT system by exploiting the natural dynamics of speckle decorrelations rather than trying to overcome them. We achieve this by adapting the principles of super-resolution optical fluctuations imaging (SOFI), originally developed for imaging blinking fluorophores, to AOT. We show that naturally fluctuating optical speckle grains can serve as the analogues of blinking fluorophores, enabling super-resolution by statistical analysis of fluctuating acousto-optic signals. 
\end{abstract}

\begin{document}

\maketitle 

\section{Introduction}

Optical microscopy is an indispensable tool in biomedical investigation and clinical diagnostics. However, its penetration depth is limited to a fraction of a millimeter inside complex samples, such as biological tissue, due to light scattering. While non-optical imaging techniques, such as those based on ultrasound sonography or magnetic resonance imaging (MRI) allow deeper investigations, their resolution is typically orders of magnitude inferior to that of optical microscopes. As a result, it is currently impossible to conduct microscopic investigations at depths, an important goal that is at the focus of many recent works \cite{ntziachristos2010going}.

The state-of-the-art approaches for deep-tissue high-resolution optical-contrast imaging are based on the combination of light and sound \cite{ntziachristos2010going}. These techniques combine the advantages of optical contrast with those of the near scatter-free propagation of ultrasound in soft tissues. The leading deep-tissue imaging techniques can be divided to two: acousto-optic tomography (AOT) \cite{elson2011ultrasound,wang2004ultrasound,resink2012state}, and photoacoustic tomography (PAT) \cite{wang2004ultrasound,wang2012photoacoustic}. PAT relies on the generation of ultrasonic waves by absorption of light in a target structure under pulsed optical illumination. In PAT, images of absorbing structures are reconstructed by recording the propagated ultrasonic waves with detectors placed outside the sample. In contrast to PAT, AOT does not require optical absorption, but is based on the acousto-optic effect: in AOT a focused ultrasound spot is used to locally modulate light at chosen positions inside the sample. The ultrasound spot is generated and scanned inside the sample by an external ultrasound transducer, as shown in Fig. 1(a). The modulated, frequency-shifted light, is detected outside the sample by one of a variety of interferometry-based approaches \cite{elson2011ultrasound,resink2012state}. This enables the reconstruction of the light intensity traversing through the localized acoustic focus inside the sample. 
AOT and PAT thus both provide images of optical contrast with a spatial resolution limited by the dimensions of the ultrasound focus, which is dictated by acoustic diffraction. Ultimately, the ultrasound focus size in soft tissue is limited by the attenuation of high-frequency ultrasonic waves. As a result, the practical attainable resolution deteriorates with imaging depth. Typically providing a depth to resolution ratio of approximately 100 \cite{wang2012photoacoustic,beard2011biomedical}. 

Since the acoustic-scale resolution of AOT and PAT does not allow microscopic imaging, many efforts have been devoted to develop approaches that can surpass the acoustic resolution limit \cite{horstmeyer2015guidestar}. 
In PAT, nonlinear effects \cite{lai2015photoacoustically}, dynamic speckle illumination \cite{chaigne2016super,hojman2017photoacoustic}, temporal fluctuations from flowing absorbers \cite{chaigne2017super}, or localization of isolated absorbers \cite{vilov2017overcoming,dean2018localization} were exploited to provide sub-acoustic resolution. However, these photo-acoustic based approaches still require optical absorption and relative intense laser pulses. In AOT, surpassing the acoustic resolution limit was first demonstrated using nonlinear acousto-optic effects \cite{selb2002nonlinear}, requiring high acoustic-pressures. Recent approaches that rely on optical wavefront-shaping to focus light into the acoustic focus can surpass the acoustic diffraction limit via either digital phase-conjugation \cite{si2012breaking,judkewitz2013speckle}, or by measurement of the acousto-optic transmission matrix \cite{katz2017controlling}. Unfortunately, these novel wavefront-shaping based approaches require the performance of a very large number of measurements (up to several thousands \cite{katz2017controlling,judkewitz2013speckle}) \textit{and} digital wavefront-shaped illumination, all within the speckle decorrelation time of the sample.

\begin{figure*}[t!]
\centering
\includegraphics[width=0.76\textwidth]{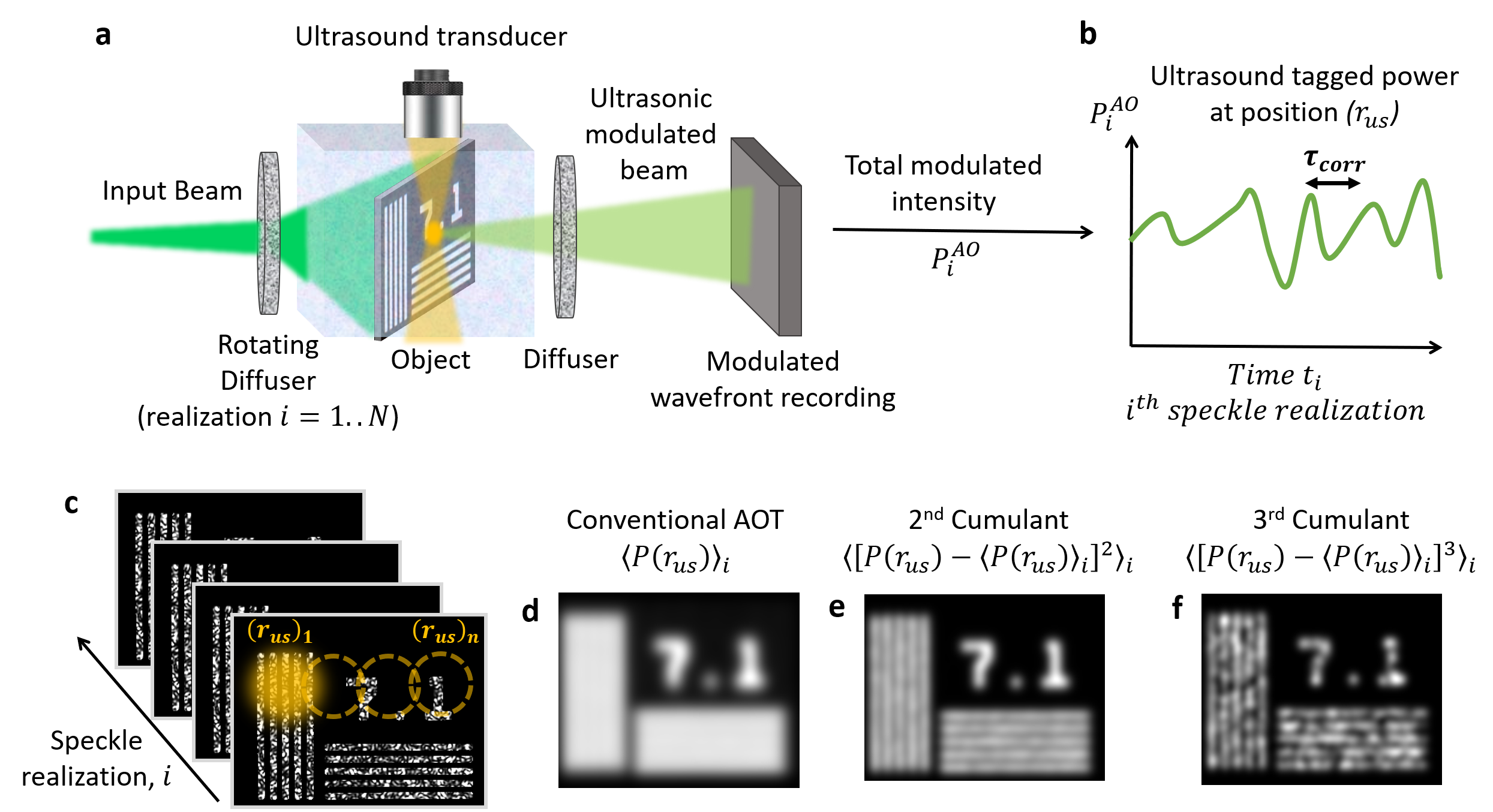}
\caption{Principle of super-resolution acousto-optic tomography (AOT) via speckle fluctuations and numerical results: (a) Schematic of the experimental setup: A conventional AOT setup is equipped with a rotating diffuser for producing controlled speckle fluctuations. An object hidden inside a scattering sample is imaged  by scanning a focused ultrasound beam (in orange) over the object, and recording the ultrasound-modulated light power at each position. (b) At each position of the ultrasound focus, the fluctuations of the ultrasonically-modulated light power due to the diffuser rotation (speckle decorrelation) are recorded. These are statistically analyzed to provide super-resolved AOT images (c-f). (c) Schematic depiction of the source of AOT fluctuations: at each position of the ultrasound focus (orange), different speckle realizations lead to different ultrasound-modulated power. (d) An image produced with the average detected signal at each position provides a conventional AOT image resolution. (e) An image produced from the second-order cumulant (variance) of the fluctuations at each position provides a $\sqrt{2}$  improved resolution withtout deconvolution. (f) The third-order cumulant image provides a $\sqrt{3}$  improved resolution without deconvolution.
}
\label{fig:false-color}
\end{figure*}

As the speckle decorrelation time in tissue can reach sub-millisecond timescales \cite{liu2015optical}, these approaches for super-resolution AOT have so far been limited to static samples.
Here, we present a novel approach for super-resolution AOT, which, unlike previous works, \textit{exploits} the naturally short speckle decorrelation time of dynamic samples rather than trying to overcome it. Our approach is made possible by adapting the principles of super-resolution optical fluctuations imaging (SOFI) \cite{dertinger2009fast}, a super-resolution technique originally developed for microscopic imaging of blinking fluorophores, to AOT. Specifically, we show that naturally dynamic fluctuating optical speckle grains in AOT can be considered as the analogues of blinking fluorescence molecules of SOFI, allowing super-resolved AOT. We show that a super-resolved AOT image can be obtained by statistical analysis of temporal fluctuations of ultrasonically modulated light, recorded using a conventional AOT setup. 
Our work extends the application of the fundamental principles of SOFI 
from microscopy \cite{dertinger2009fast}, ultrasound sonography \cite{bar2017fast}, and photoacoustics \cite{chaigne2016super,chaigne2017super}, to AOT.

Most importantly, since in our approach super-resolution originates from the natural temporal fluctuations of speckles, only a single measurement needs to be performed within the speckle decorrelation time, in contrast to the thousands of measurements (and wavefront-shaping processes) required by the state of the art approaches \cite{judkewitz2013speckle,katz2017controlling}. Our approach thus allows to perform super-resolved AOT with speckle decorrelation times orders of magnitude shorter than current approaches. 

\section{Principle}
\label{sec:examples}

The principle of our approach is presented in Fig. 1. In our approach, a set of measurements performed using a conventional AOT setup is used to obtain super-resolution. Figure 1(a) shows a schematic description of a typical AOT experiment, where the goal is to image a target object hidden inside a scattering sample. In AOT, a laser beam at a frequency $f_{opt}$ illuminates the sample, and an ultrasound wave at a central frequency $f_{US}$ is focused at a position, $r_{us}$, which is scanned across the volume of interest. The acousto-optic signal for each ultrasound focus position, $P^{AO}(\mathbf{r_{us}})$, is the total detected power of the light that has been frequency-shifted to $f_{AO}=f_{opt}+f_{US}$, by the acousto-optical interaction at the focus. Considering linear ultrasound modulation and an ultrasound focus pressure distribution given by $h(r-r_{us})$, the acoustically-modulated power, as measured by an ideal detector placed outside the sample (see Discussion) is given by:

\begin{equation}
   P^{AO}(\mathbf{r_{us}})=\iint h(\mathbf{r}-\mathbf{r_{us}})I_{obj}(\mathbf{r})dv=[h\ast I_{obj}](\mathbf{r_{us}})
\label{eq:refname1}
\end{equation}
Where the integration is performed over the entire sample volume. Equation 1 shows that the AOT image in the ideal case, $P^{AO}(r_{us})$, is the convolution of the optical intensity distribution inside the medium, $I_{obj}(r)$, with the ultrasound focus pressure distribution, $h(r)$. Thus, the imaging resolution in AOT is dictated by the ultrasound focus size, which serves as the effective PSF of the AOT system. 

As the common laser illumination in AOT is spatially and temporally coherent, the light intensity distribution inside the medium, $I_{obj}(\mathbf{r})$, is given by the product of the target object transmission, $T_{obj}(\mathbf{r})$ with an illumination speckle intensity pattern, $S(\mathbf{r})$: $I_{obj}(\mathbf{r})=T_{obj}(\mathbf{r})S(\mathbf{r})$. Due to the natural dynamics of the sample, the speckle pattern illuminating the object $S(\mathbf{r})$ dynamically fluctuates at timescales given by the speckle decorrelation-time, $\tau_{corr}$. The speckle decorrelation time can reach sub-millisecond timescales in living tissue \cite{liu2015optical}, forming one of the fundamental challenges for wavefront-shaping based imaging techniques \cite{horstmeyer2015guidestar}. However, as we show below, the temporally fluctuating speckles can be utilized rather than struggled with. 

Defining the speckle pattern illuminating the object at time $t_{i}$ by $S_{i}(\mathbf{r})$, the measured ultrasonically-tagged intensity at the time $t_{i}$ is given by:

\begin{equation}
P^{AO}(\mathbf{r_{us}},t_{i})\equiv\mathbf{P}_{i}^{AO}(\mathbf{r_{us}})=[h\ast(T_{obj}\times S_{i})](\mathbf{r_{us}})
\label{eq:refname2}
\end{equation}
Equation (2) shows that the conventional AOT intensity measurements, $P^{AO}(\mathbf{r_{us}},t_{i} )$, are expected to temporally fluctuate at timescales given by $\tau_{corr}$, as shown in Figure 1(b). Most importantly, Equation 2 shows that the measured intensity fluctuations in AOT are mathematically equivalent to the measured intensity fluctuations in fluorescence microscopy with blinking fluorophores, utilized in SOFI\cite{dertinger2009fast}: the fluctuating speckle grains, $S_i$, effectively serve as blinking fluorophores on an object having the spatial labeling distribution  $T_{obj}(\mathbf{r})$, which is imaged by a microscope with a PSF given by $h(\mathbf{r})$. 

Since the temporal fluctuations of individual speckle grains are uncorrelated \cite{goodman2015statistical}, the principles of SOFI can be directly applied to AOT. Thus, all that is required for super-resolved AOT is to acquire a set of  $i=1:m$ conventional AOT measurements, temporally separated by more than $\tau_{corr}$, at each position of the ultrasound focus.  In case that $\tau_{corr}$ is too long, a rotating diffuser can be introduced at the illumination path, for generating rapid fluctuations at controlled timescales, as shown in Fig. 1(a-b). To achieve super-resolution via SOFI, for each probed position $\mathbf{r_{us}}$ (each AOT image 'pixel'), the $n^{th}$-order statistical cumulant, $C_n(\mathbf{r_{us}})$, of the recorded temporal intensity fluctuations is calculated and taken as the reconstructed pixel intensity for $\mathbf{r_{us}}$. The $n^{th}$-order cumulant provides a $\sqrt{n}$-times resolution increase without deconvolution (see Supplementary section 2), and up to $n$-times  resolution increase with deconvolution \cite{dertinger2009fast}. For example, the second-order cumulant:
\begin{equation}
C_{2}(\mathbf{r_{us}})=\left\langle \left[P_{i}^{AO}(\mathbf{r_{us}})-\left\langle P_{i}^{AO}(\mathbf{r_{us}})\right\rangle _{i}\right]^{2}\right\rangle _{i}
\label{eq:refname3}
\end{equation}
which is simply the variance, provides a $\sqrt{2}$ resolution increase before deconvolution, and a factor of 2 resolution increase with deconvolution. For SOFI to work, all that is required is the presence of uncorrelated temporal fluctuations of sub acoustic-diffraction sized sources of signals. A condition which is naturally fulfilled by randomly fluctuating speckles (See details in Supplementary Section S2).

Figure 1(c) depicts the proposed data acquisition scheme utilizing a conventional AOT system, along with numerically simulated results using $m=2,000$ speckle realizations: the acoustic focus (in orange) scans the object (a USAF resolution target). For each acoustic focus position the frequency-shifted light intensity is measured repeatedly at times $t_1..t_m$, with a different speckle pattern illuminating the object at each time. While the conventional AOT image, or the average of the measured intensities at each point, provides an image blurred by the acoustic PSF, as depicted in Fig. 1(d), the higher-order cumulants images shown in Figs 1(e-f) allow to resolve the target features beyond the resolution of the ultrasound focus. 

\section{Results}
To experimentally demonstrate our approach, we constructed a proof-of-principle experiment using the setup schematically shown in Fig. 1(a), and explained in detail in Supplementary Fig. S1. The setup is a conventional pulsed AOT setup, with the only addition of a controlled rotating diffuser before the sample, for generating random speckle pattern with controlled decorrelation time.  
The ultrasound focus was provided by an ultrasound transducer (V315, Olympus, F\#=1.33), driven by $300ns$ long sinusoidal pulses at $f_{us}=10MHz $ central frequency, with a $150Vpp$ amplitude.
The ultrasound focus position was horizontally scanned using a motorized translation stage (Thorlabs), and its vertically position was controlled by adjusting the relative delay between the ultrasound pulses and the optical pulses from the q-switched laser (Standa STA-01, providing 1ns long pulses, at 25KHz repetition rate, 532nm wavelength). The ultrasound-modulated light was detected via off-axis phase-shifting digital holography \cite{atlan2005pulsed}, using a high-resolution sCMOS camera (Zyla 4.2 Plus, Andor), and a frequency-shifted reference arm.
To allow direct optical inspection of the target object and ultrasound focus, we used a sample composed of a target object (A portion of a USAF 1963A resolution target) placed in a transparent water tank between two scattering diffusers. 

A direct optical image of the object illuminated by one speckle realization, as captured by removing the second diffuser and mounting an imaging lens, is shown in Fig. 2(a). We used the same configuration to characterize the ultrasound focus (see Supplementary Fig. S2), and the speckle grain size. The dimensions of the ultrasound focus were {$\Delta X=350\mu m , \Delta  Y=350\mu m$ } full-width at half max (FWHM) in the horizontal and vertical directions, correspondingly. 

To demonstrate our approach we performed a conventional AOT with the second diffuser in place, by scanning the ultrasound focus over the path marked by a dashed orange line in Fig. 2(a). Different from conventional AOT, for each ultrasound focus position, $\mathbf{r_{us}}$, we recorded $m=160$ different ultrasound-modulated light intensities, ${P}_{i}^{AO}(\mathbf{r_{us}})$, each with a different (unknown) speckle realization, $S_{i}(\mathbf{r})$ (i=1..m), obtained by rotating the controlled diffuser placed before the sample. 

\begin{figure}[t!]
\includegraphics[width=\linewidth]{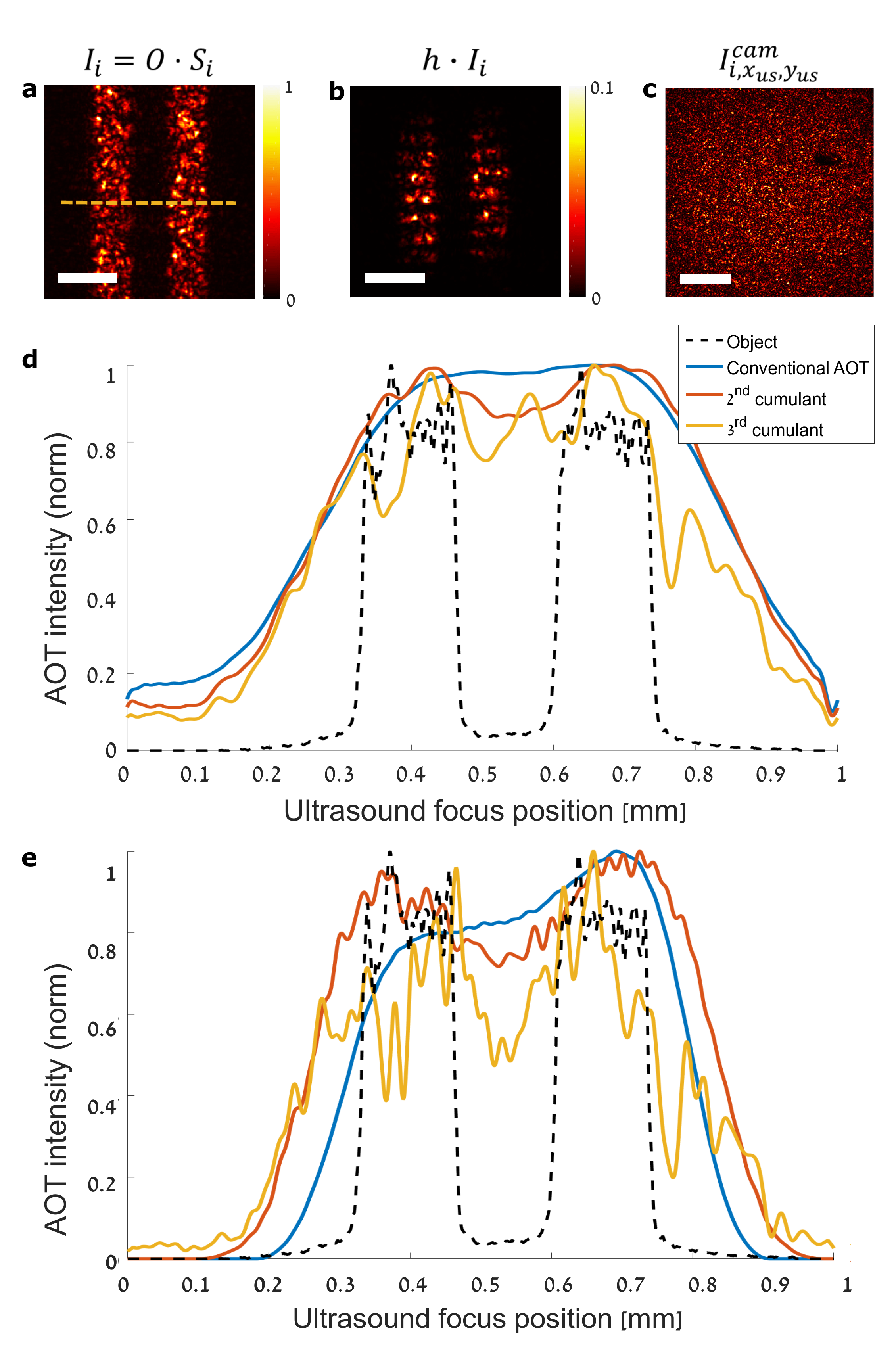}
\caption{Experimental demonstration of super-resolution AOT using 160 speckle realizations, with a large speckle grain size: (a) target object, I(x,y) (a section of USAF1963A resolution target), illuminated by a single speckle realization, $S_{i}(x,y)$, as imaged directly without the scattering layer. (b) ultrasonically-tagged optical intensity distribution at the object plane, for one position ($x_{us},y_{us}$) of the ultrasound focus, $h(x,y)$, visualizing the ultrasound focus size. (c) Detected ultrasound-modulated light pattern for (b), as recorded at the camera plane through the second diffuser. (d) AOT image profile obtained by scanning the ultrasound focus along the dashed line in (a):  blue - conventional AOT resolution trace, obtained from the ensemble average of the recorded power at each position; red - square-root of the second-order cumulant of the acousto-optic fluctuations, showing improved resolution; orange - third order cumulant; black, dashed: cross-section of the target object, along the dashed line in (a). (e) The results from (d) after deconvolution. scale-bars: 200$\mu$m (a,b), 330$\mu$m (c). }
\label{fig:false-color}
\end{figure}

The experimental results obtained with this setup for two different speckle grain size are presented in Figures 2-3. 
An example for the raw detected acoustically-modulated light pattern at the camera plane, for one ultrasound position $\mathbf{r_{us}}$ and speckle realization $i$, which is a random speckle pattern, is shown in Figure 2(c). For each measurement, the total acoustically-modulated power, ${P}_{i}^{AO}(\mathbf{r_{us}})$,  is calculated from this pattern by integration over all camera pixels. To demonstrate super-resolution AOT, the first three statistical cumulants of ${P}_{i}^{AO}(\mathbf{r_{us}})$ were calculated for each ultrasound focus position $\mathbf{r_{us}}$. 
Figure 2(d) shows the one-dimensional AOT traces obtained using our proposed approach. While the first cumulant (realizations average), having the conventional AOT resolution, does not resolve the target features, the higher order cumulants clearly resolve the two target lines. To display a fair comparison between the different cumulants orders, each of the plotted traces is the $N^{th}$-root of the $N^{th}$-order cumulant.
As in conventional SOFI, additional resolution improvement, up to a factor of $N$ for the $N^{th}$-order cumulant, can be obtained by deconvolving the $N^{th}$-order cumulant trace with the $N^{th}$ power of the measured acoustic PSF, $h(t)$ \cite{dertinger2010achieving}. Figure 2(e) presents the results of a such a deconvolution, performed via Richardson-Lucy deconvolution on the experimentally measured cumulant traces. The resolution increase in the high order cumulants is enhanced in the deconvolved traces, as expected. 

Different from conventional SOFI in fluorescence microscopy \cite{dertinger2009fast}, where the labelling concentration of the fluorescent molecules is controlled via sample preparation, in AOT the number of fluctuating speckle grains contained inside the acoustic focus, $N_{speckles}$, is determined by the ratio between the ultrasound focus size, $D_{US}$, and the size of the optical speckle grains, $D_{speckle}$: $N_{speckles} \approx (D_{US}/D_{speckle})^2$. The larger is the number of  uncorrelated fluctuating speckles contributing to the AOT signal, the smaller is the relative amplitude of fluctuations compared to the mean measured AOT signal. Quantitatively, the AOT signal standard-deviation is $\sqrt{N_{speckles}}$ times smaller than the mean AOT signal. Thus, for given experimental signal-to-noise conditions, applying SOFI to AOT is expected to be more challenging the larger is $N_{speckles}$, as shown in Supp. Figure S4. However, more importantly, $N_{speckles}$ affects also the statistical estimation accuracy of the cumulants, even in the absence of measurement noise, as we explain below.

The results of Fig.2 were obtained with an optical speckle grain size of  $\sim 8 \mu m$ at the target plane (HWHM of the speckle spatial autocorrelation, as imaged by a camera), yielding \textbf{ $N_{speckles}\approx 600$ }. The speckle grain size in this first experiment was set by controlling the beam diameter on the rotating diffuser and removing the first diffuser in the sample. 
While the speckle grains in this experiment are considerably smaller than the ultrasound focus, they are still considerably larger than the speckle grain size expected inside volumetric scattering samples such as tissue, where the speckle grain size is of the optical wavelength scale. 
Considering high frequency $(>50MHz)$ ultrasound, and infrared laser illumination,  $N_{speckles} > \sim2,500$ speckle grains are expected to be contained within the ultrasound focus in deep-tissue AOT experiments.

To demonstrate our approach with smaller speckle grains, we have repeated the experiments of Fig.2 with a speckle grain size of  $\sim 2.45 \mu m$, by mounting all three diffusers in place. This speckle grain size results in  \textbf{ $N_{speckles}\approx 4,800$} over the entire area of the ultrasound focus. To minimize the effective number of contributing speckle grains we have chosen to use laser pulses having a temporal duration that is significantly shorter than $1/f_{us}$. Such a choice results in an acousto-optic PSF that is axially-modulated by the ultrasound wavelength, as can be seen in Figs .2b, 3b and Supplementary Figure S2a, having an effective area that is two times smaller than that of the acoustic PSF using long laser pulses \cite{judkewitz2013speckle,katz2017controlling}. The results of this experiment are shown in Fig. 3. In this experiment, while the second cumulant image resolves the two object lines with improved contrast, the third order cumulant is corrupted by strong artifacts.

\begin{figure}[t]
\includegraphics[width=\linewidth]{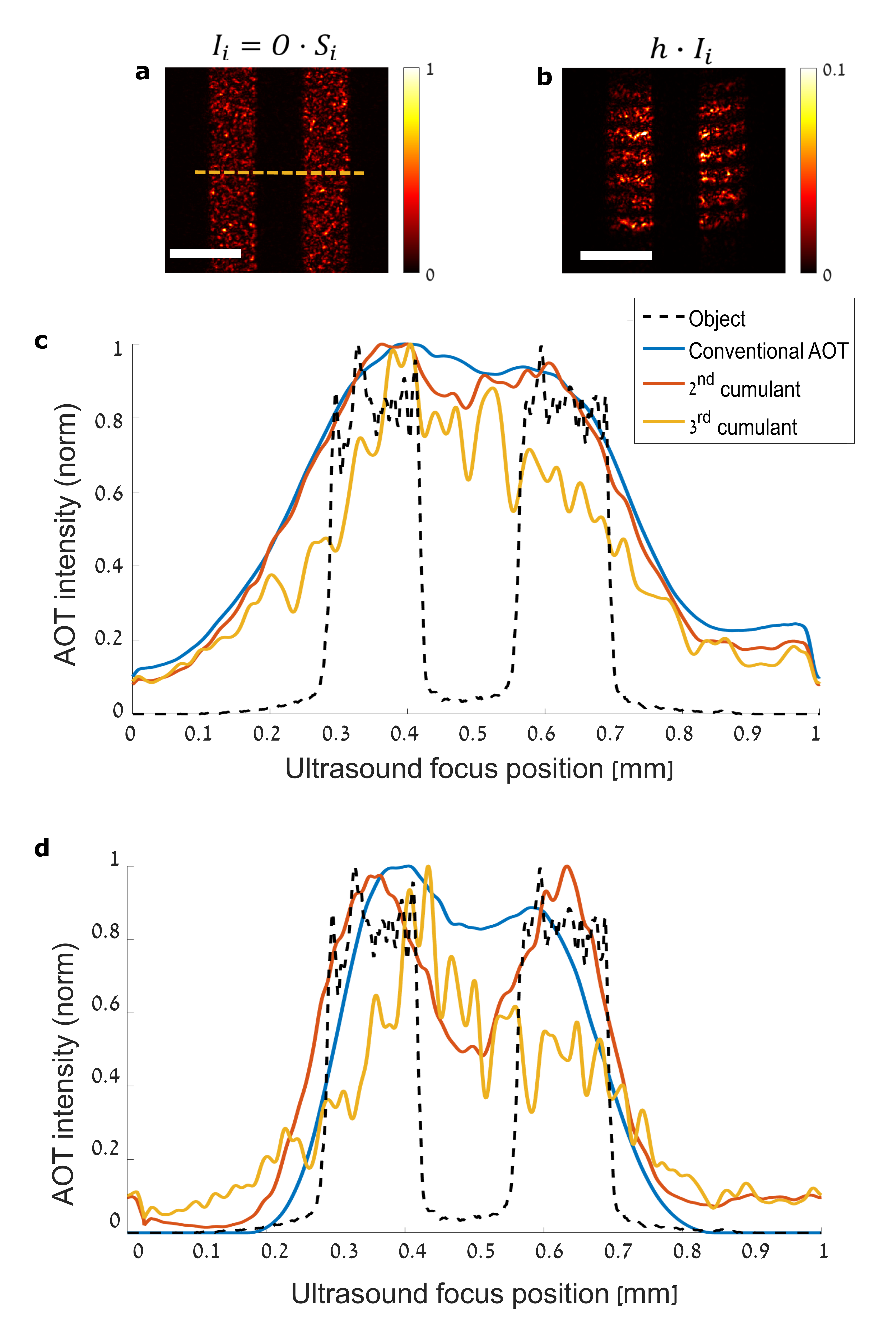}
    \caption{Super-resolution AOT with small speckle grains: (a) direct image of the target object, as illuminated by a single speckle realization. (b) Ultrasonically-tagged optical intensity distribution at the object plane, for one position of the ultrasound focus, $h(x,y)$. (c) AOT profiles along the dashed line in (a):  blue - conventional resolution AOT (average measured power); red - square-root of the second-order cumulant of the acousto-optic fluctuations, showing improved resolution; orange - third order cumulant, suffering from statistical estimation errors (see Fig.4); black, dashed: cross-section of the target object.(d) The results from (c) after deconvolution., scale-bars 200$\mu$m}
\label{fig:false-color}
\end{figure}

Interestingly, and of high practical importance, the increased noise in the third (and higher-order) cumulants is not a result of experimental conditions, but represents a fundamental limitation of SOFI, originating from the very nature of the cumulant statistical analysis \cite{muller2004cumulant}. While SOFI principle promises a potential infinite resolution increase for increasing cumulants order, such an increase is only possible for an \textit{infinite} number of speckle realizations (ensemble averaging), even in the absence of measurement noise. For a finite number of realizations (finite sample size), the inherent statistical estimation errors of higher-order cumulants \cite{muller2004cumulant} present the dominant source of image artifacts in SOFI. 

To quantitatively study the theoretical and practical limitations of high order cumulants estimation on our approach, as a function of $N_{speckles}$, and number of realizations, we have performed a set of numerical simulations in the noise-free case. The results of this study are presented in Fig. 4 and Supplementary Fig. S3. Additional results that include the effect of measurement noise are given in Supplementary Figure S4. Details on the simulations are given in Supplementary Sections 3-4.
Fig. 4(a) presents the relative estimation errors of the $N$-th root of the first three cumulants as a function of $N_{speckles}$, for m=200 and m=2,000 realizations. The estimation error of the third-order cumulant grows as a function of $N_{speckles}$, while the estimation errors of the second cumulant (the standard deviation) is largely insensitive to $N_{speckles}$. The effect of the estimation errors on simulated AOT traces, in the absence of measurement noise, is presented in Fig. 4(b-c). While the effect is small for a small number of speckles ($N_{speckles}=6$, Fig. 4(b)), it is dominant for a large number of speckle ($N_{speckles}=4000$, Fig. 4(c)). 
Thus, obtaining high quality AOT images using high-order cumulants can only be achieved in practical AOT conditions by averaging a large number of measurements, or when a small number of speckles are transmitted through the target objects, i.e. for sparsely transmitting (mostly absorbing) objects. 

The inherent difficulty in accurately estimating the high-order statistical cumulants for large $N_{speckles}$ can be intuitively understood from the central limit theorem: since the measured AOT signal is the sum of $N_{speckles}$ independent random variables, the larger is  $N_{speckles}$, the closer is the distribution of the AOT fluctuations in a given spatial position to a Gaussian distribution, whose cumulants of orders three and above are equal to zero.

\begin{figure}[b!]
\centering
\includegraphics[width=\linewidth]{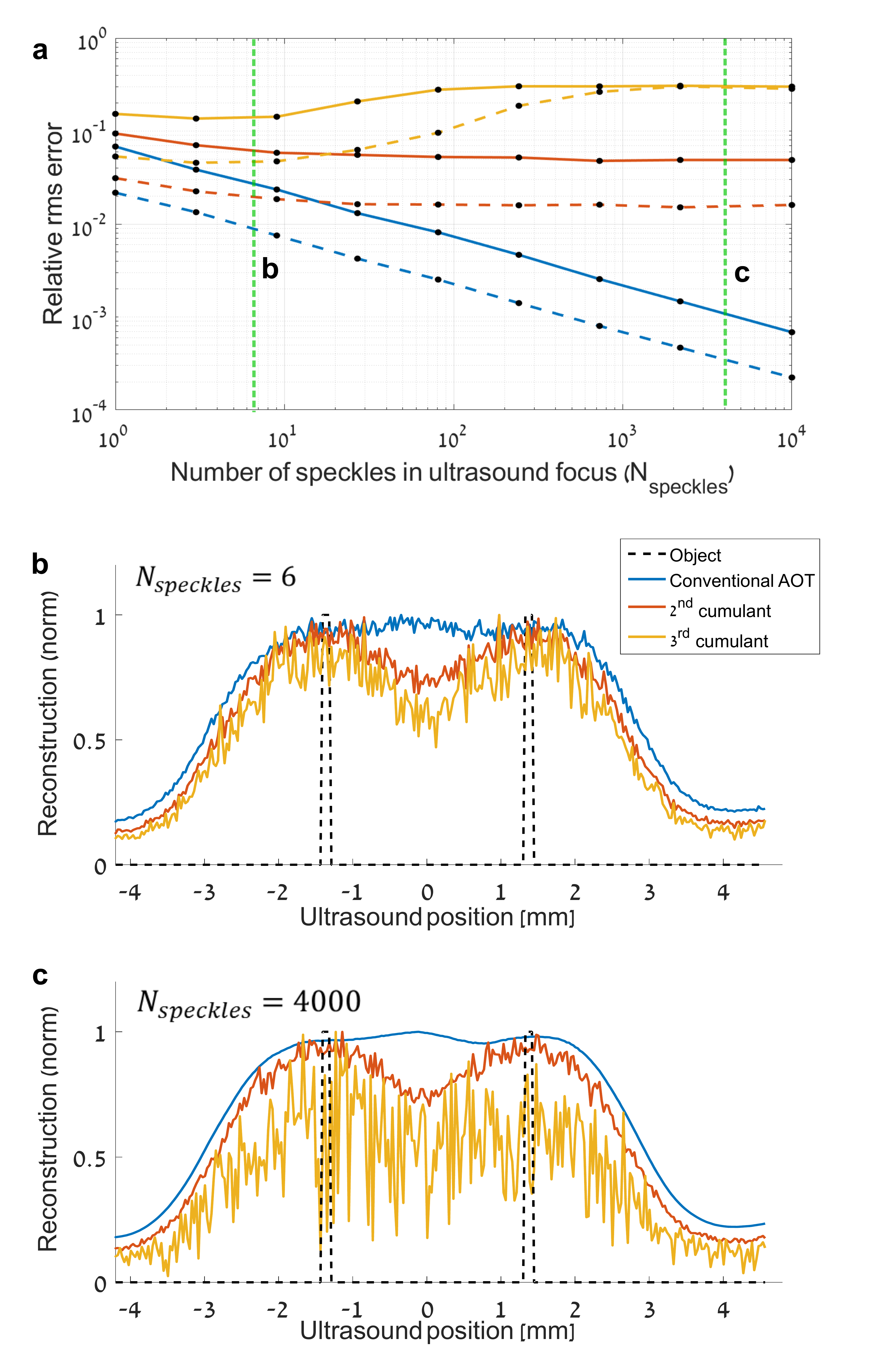}
\caption{Numerical investigation of cumulants estimation error for noise-free measurements: (a) Relative estimation error of the $N^{th}$ root of the first three cumulants as a function of the number of speckles, $N_{speckles}$, contained inside the acoustic focus, for 200 realizations (continuous lines) and 2,000 realizations (dashed lines). blue-first cumulant, red-second cumulant, orange-third cumulant. (b-c) Simulated AOT reconstructions for noiseless measurements with 400 realizations, for two different values of $N_{speckles}$, marked by the dashed green lines in (a): (b) $N_{speckles}=6$, and (c) $N_{speckles}=4,000$. The larger estimation errors of higher order cumulants in (c) are in-line with the experimental results of Fig.3c.}
\label{fig:false-color}
\end{figure}

Another, more conventional, factor that affects the estimation accuracy is measurement noise. For the approach to work, the fluctuations amplitude has to be larger than the measurement noise. We study the effect of measurement noise numerically in Supplementary Fig. S4. A larger number of realizations, $m$, can be used to mitigate the effects of both estimation errors and measurement noise, sacrificing acquisition time. 

\section{Discussion}
 
We have presented proof of principle results for super-resolved AOT using dynamic speckle fluctuations, and have analyzed the sensitivity of the approach to measurement noise and the inherent estimation errors of finite statistics.  

Another important factor that affects the effective measurement accuracy in AOT is the number of optical modes (speckles) that are detected by the AOT system.  Equation (1) implicitly assumes ideal acousto-optic detection, i.e. an AOT detector that detects \textit{all} of the acoustically-modulated light power, i.e. collects all the optical modes. However, in practice, any optical detector has a limited etendue, which can be characterized by the number of optical speckle grains, $N_{det}$, that can be detected by it. For the digital-camera based system used in our experiments, the number of camera pixels poses the limit on the number of speckle grains that can be detected with a high signal to noise (SNR). 
In deep tissue imaging, the number of total speckles in the scattered field outside the sample may be considerably larger than $N_{det}$. In such conditions, the relative error of the total measured intensity, $P^{AO}$, only due to speckle statistics is: $\epsilon = \Delta {P^{AO}}/P^{AO} \approx 1/\sqrt{N_{det}}$. This error is only due to speckle statistics, and is added to any other noise sources. In order to be able to accurately measure small fluctuations, a larger number of speckles needs to be detected at the camera plane. For example, for the second SOFI cumulant (i.e. the variance), the ratio of the square root of the second cumulant to mean measured intensity is: $\sqrt{C_2}/C_1 = \sqrt{C_2}/P^{AO} \approx 1/\sqrt{N_{speckles}}$. Thus the SNR for measuring the square root of the second cumulant in the case of no measurement noise is: $SNR=\sqrt{C_2}/\Delta {P^{AO}}={\sqrt{N_{det}/N_{speckles}}}$. To allow super-resolved AOT using our approach, the experimental AOT system should provide $N_{det}\gg N_{speckles}$. This condition is readily met with modern megapixel-count cameras, or nonlinear-crystals based detection \cite{laudereau2016ultrafast}.

Our approach is general, and can be applied to any conventional AOT system. It does not necessitates wavefront-measurements, SLMs, complex computations, non-linear effects, or memory-effect speckle correlations \cite{katz2014non}, as required by alternative approaches. Most importantly, our approach is designed to work with short speckle decorrelation times, which is one of the major limiting factors of current approaches \cite{katz2017controlling,judkewitz2013speckle,si2012breaking}. 

While our approach relieves the requirements on the speckle decorrelation time, it relies on multiple measurements at each position of the ultrasound focus, thus increasing the acquisition time, a fundamental quality shared with many super-resolution imaging approaches. 
In our proof-of-principle experiments, we have performed 160 measurements at each spatial position, using a setup that was not optimized for speed. Thus the current acquisition time, composed of 24 phase-shifting images per measurement, was 2 seconds per realization, excluding diffuser rotation and data processing. The acquisition time can be shortened by orders of magnitude using faster detection approaches, such as lock-in camera detection \cite{liu2016lock}, and fast cameras \cite{liu2017focusing}.  
Another approach for improved acquisition time is using an ultra-fast plane-wave AOT approach \cite{laudereau2016ultrafast}, based on nonlinear crystals.
  
We have demonstrated the use of basic cumulants for super-resolution AOT. Additional statistical analysis techniques, such as cross-cumulants analysis \cite{dertinger2010achieving}, and balanced SOFI\cite{geissbuehler2012mapping}, could further improve the resolution and image quality.
Other, more advanced algorithms, such as compressed-sensing sparsity-based reconstruction, should improve the resolution and reconstruction fidelity even further, as was recently demonstrated in PAT \cite{hojman2017photoacoustic,murray2017super}. 

We used a controllable diffuser to produce controlled speckle decorrelation. This may not be required in \textit{in-vivo} imaging, where the natural speckle decorrelation caused by blood flow or tissue decorrelation may provide the source for fluctuations \cite{chaigne2017super}, turning the natural sample dynamics into a positive effect in AOT. 

\section*{Funding Information}
European Research Council (ERC) Horizon 2020 research and innovation program (grant no. 677909), Azrieli foundation, Israel Science Foundation.

\section*{Acknowledgments}
We thank Prof. Hagai Eisensberg for the q-switched laser.

\bibliography{SOFI_paper}

\clearpage



\section*{Supplementary material}

# Acousto-optic tomography beyond the acoustic diffraction-limit using speckle decorrelation: Supplementary Information

**DANIEL DOKTOFSKY**[1], **MORIYA ROSENFELD**[1], **AND ORI KATZ**[1,*]

[1] *Applied Physics Department - Hebrew University of Jerusalem, Israel*
[*] *Corresponding author: orik@mail.huji.ac.il*

**This document provides supplementary information to "Acousto-optic tomography beyond the acoustic diffraction-limit using speckle decorrelation". It provides a detailed description of the experimental setup, the theoretical framework for resolution increase via cumulant analysis, an analytic estimation of the fundamental signal to noise (SNR) limitations of acousto-optic tomography (AOT) measurements, and a numerical study of the effects of measurements noise on super-resolved image reconstruction.**

## 1. EXPERIMENTAL SETUP

The experimental setup used in this work is shown schematically in Fig. S1. It is a conventional pulsed acousto-optic tomography (AOT) experiment, employing a camera-based digital phase-shifting holography detection [1]. The light source was a single longitudinal-mode 532nm passive Q-switched laser (Standa STA-0 ) providing $< 1ns$ duration pulses at $f_{rep} = 25kHz$ repetition rate. The laser output was split to a reference and object arm by a polarization beam splitter (PBS), with a half-wave plate (HWP) controlling the splitting ratio. At the object arm, the beam width was expanded to approximately 7mm and was introduced into the tank through a diffuser (Newport 1.5$^o$ light shaping diffuser) mounted on a computer-controlled rotation stage, and an additional diffuser placed inside the water tank (Newport 5$^o$ light shaping diffuser) generating a speckled beam at the target plane, with controlled speckle decorrelation time. As a target object we have used two stripes from group 3.6 of a USAF 1963A target (Thorlabs), placed 5cm behind the second diffuser. An Olympus v315 ultrasound transducer provided the ultrasound pulses for acousto-optic tagging. The transducer was driven by sinusoidal pulses at $f_{us}$=10MHz frequency, 300$ns$ duration (rectangular envelope), and $150Vpp$ amplitude, provided by function generator (Keysight 33600A) connected to an RF amplifier (Amplifier Research 25A250A). The ultrasound transducer was mounted on a computer-controlled motorized translation stage (Thorlabs), which allowed horizontal scanning of the acoustic focus. The vertical position of the ultrasound focus was controlled by varying the relative delay between the acoustic pulses and the laser trigger pulses. To ensure accurate temporal synchronization of the $< 1ns$ laser pulses, which allowed to 'freeze' the ultrasound focus motion and resolve the acoustic wavelength fringes inside the acoustic focus (shown in Supplementary Figure S2), and compensate for *slow* temporal drifts of the laser pulses with respect to the laser trigger, a photodiode was used to trigger the acoustic pulse from the optical

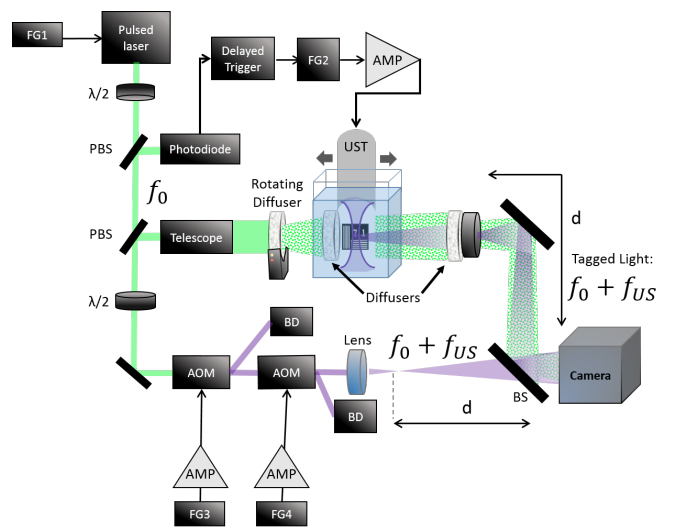

**Fig. S1.** The experimental setup. PBS - polarized beam splitter. AOM - acousto optic modulator. BD - beam dump. FG - function generator. AMP - amplifier. UST - ultrasonic transducer. $f_0$ - original optical frequency. $f_{US}$ - ultrasound frequency.

output of the laser (synchronizing the acoustic pulse with the next laser pulse) . The modulated light after the target object passed another diffuser (Newport $5^o$ light shaping diffuser) and a controllable iris before reaching the camera (Andor Zyla 4.2 Plus sCMOS camera).

At the reference arm, two acousto-optic modulators (AA Opto-Electronic MT80-A1) connected to RF amplifiers (Mini-Circuits ZHL-3a) were used to frequency shift the reference by $10MHz + 12.5KHz + 5Hz$. The reference arm was thus shifted by $12.5KHz + 5Hz$ relative to the ultrasound modulated photons in order to perform phase-shifting holography, using the sCMOS camera operated at a frame-rate 20 frames per second [2, 3], with efficient rejection of the coherent interference background of the unmodulated light, originating from the use of the short laser pulses (compared with the ultrasound period and ultrasound pulse width). Efficient background rejection of the unmodulated light was obtained by adding a frequency-shift of $f_{rep}/2 = 12.5KHz$ [2], to the phase-shifting $5Hz$ frequency-shift. This rejection can be understood in terms of sampling: the laser repetition rate effectively samples at $25KHz$ the intensity of the interference between the reference arm and object arm, thus unmodulated light having a frequency shift of $12.5KHz$ will be 'sampled' at alternating phase of $[0, \pi]$ between two sequential pulses, averaging the interference result to zero, and thus effetively filtering the unmodulated light background interference.

For the AOT measurements, horizontal scans of the ultrasound focus were performed at $10\mu m$ steps using the motorized translation stage. At each step measurements of the ultrasound modulated signal were performed for 160 different realizations of the diffuser. The cumulants were then calculated for each position, as explained in Supplementary Section 2 below. The resulting spatial AOT scan of each cumulant was smoothed by a convolution with a Gaussian kernel having a standard-deviation of 1.125 spatial pixels (equivalent to $11.25\mu m$).

A direct optical characterization of the ultrasound modulating focus, as measured by removing the last diffuser and mounting an imaging lens, and averaging over 200 speckle realizations, is presented in Figure S2(a,b). The short laser pulses used allow to clearly resolve the ultrasound wavelength vertical fringes inside the acoustic focus (effectively 'freezing' the vertically propagating ultrasound pulse). Resolving these features, rather than having them blurred by long laser pulses, as used in most AOT systems, is valuable in order to reduce the effective area of the ultrasound focus, effectively measuring a smaller number of speckles.

In the AOT experiments, the total intensity of the modulated light recorded on the camera was calculated by summing the detected intensity over all camera pixels: for each position of the ultrasound focus $(x_{us}, y_{us})$, and j-th speckle realization, this value for a 'single pixel' in the AOT image was calculated as:

$$P_j^{AO}(x_{us}, y_{us}) = \sum_{x',y'} I_{j,x_{us},y_{us}}^{cam}(x', y') \quad j = 1,..,m \quad (S1)$$

where $I_j^{cam}(x, y)$, is the absolute value squared of the detected modulated field at the camera plane.

## 2. IMAGE RECONSTRUCTION AND RESOLUTION INCREASE USING HIGH-ORDER CUMULANTS

After measuring a set of $j = 1..m$ values of acousto-optic modulated power for ultrasound focus position, $P_j^{AO}(r_{us})$, a calculation of the $n^{th}$ order temporal auto-cumulant $C_n(r_{us})$ was

performed for each pixel in the reconstruction image. $C_n(r_{us})$ was calculated recursively using the previous-order cumulants and the $k = 1 : n^{th}$ order moments $\mu_k(r_{us}) = \left\langle \left[ P_j^{AO}(r_{us}) \right]^k \right\rangle_i$:

$$C_n(\mathbf{r_{us}}) = \mu_n(\mathbf{r_{us}}) - \sum_{k=1}^{n-1} \binom{n-1}{k-1} C_k(\mathbf{r_{us}}) \mu_{n-k}(\mathbf{r_{us}}) \quad (S2)$$

where the recursion base case is the mean image: $C_1(r_{us}) = \mu_1(r_{us})$. $C_2(r_{us})$ provides the variance image.

The $n^{th}$ cumulant yields images proportional to the convolution of the object with the $n^{th}$ power of the PSF, providing a $\sqrt{n}$ resolution increase for the case of a Gaussian PSF, without deconvolution [21]. The source of the resolution increase in SOFI is the the $n^{th}$ power of the PSF: averaging over the realizations $P_j^{AO}(x,y)$ provides the average of conventional AOT images of the object convoluted with the ultrasound focus, but does not provide increased resolution:

$$\left\langle I_j^{AO}(x,y) \right\rangle_j = \langle h(r) * [\mu_s(r) \times I(r)] \rangle \approx h(r) * I(r) \quad (S3)$$

and the resolution is dictated by the spatial frequency content of h(r). Calculating a higher ordered cumulant $C_n\{I_j^{AO}(x,y)\}_j$ where $n$ is the order of the cumulant, results in [4]:

$$C_n \left\{ I_j^{AO}(x,y) \right\}_j \propto h^n(r) * I^n(r) \quad (S4)$$

Since taking the PSF to the power of $n$ is interpreted in the Fourier domain as a $n$ convolutions of the Fourier transform of the PSF with itself, the PSF to the power of $n$ posess an $n$-times broader frequency content than the original PSF, and thus potentially provides an $n$-times resolution increase after deconvotluion. Without deconvolution, for a Gaussian-shaped PSF the resolution increase grows as $\sqrt{n}$.

## 3. CUMULANTS ESTIMATION ERROR FOR IDEAL NOISE-FREE MEASUREMENTS CASE

As mentioned in the Discussion section, and as shown in Fig. 4, the estimation errors for the $n$-th root of the $n$-th cumulants are strongly dependent on the number of fluctuating speckle grains contained inside the ultrasound focus, $N_{speckles}$, even with no measurement noise present. To complement Figure 4, in

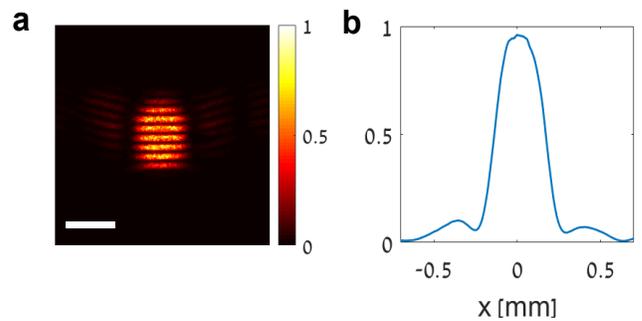

**Fig. S2.** Characterization of the acoustic focus: (a) direct image of the acoustic pulse averaged over 200 speckle realizations, clearly resolving the ultrasound wavelength oscillations. scale-bar $350\mu m$; (b) one-dimensional profile of the ultrasound focus, obtained by summing over the columns of (a), providing the effective one-dimensional PSF for deconvolution of the AOT traces.

Figure S3 we provide similar traces but for the raw cumulants (rather than their *n*-th root). As can be seen, the estimation errors increase with $N_{speckles}$ for the third order cumulant. The estimation errors for higher order cumulants are even higher (not shown).

The graphs in Figure 4 and Supplementary Figure S3 were generated by running 500 different numerical experiments for each value of speckle count, $N_{speckles}$, where in each experiment $m = 200$ or $m = 2000$ random speckle realizations (the continuous and dashed lines, correspondingly) were calculated. For each realization, the intensity sum of the $N_{speckles}$ grains was saved. Then, for each experiment the first three cumulants were calculated from the *m* recorded sum of these speckle patterns. Each point in Figure S3 was calculated by taking the ratio between the standard deviation of the cumulants values in these 500 experiments and their average values in these 500 experiments. For Figure 4, the *n*-th roots of the *n*-th cumulants were calculated, and the ratio between their standard deviation and mean was used to plot the traces in Figure 4.

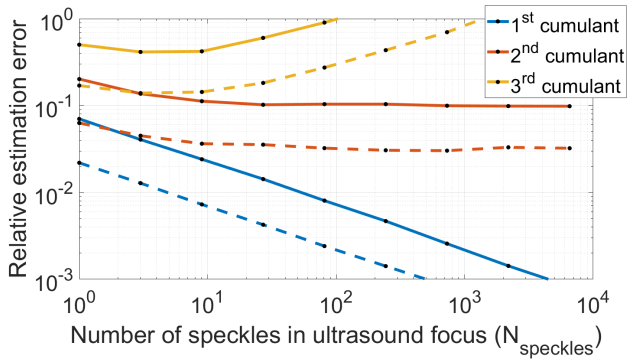

**Fig. S3.** Numerical calculation of the cumulants relative estimation errors in the ideal noise-free measuraments case, for different number of speckles in the ultrasound focus, for 200 speckle realizations (continuous line) and 2,000 realizations (dashed line).

## 4. NUMERICAL INVESTIGATION OF THE EFFECT OF MEASUREMENT NOISE

For the approach to work, the fluctuations amplitude has to be larger than the measurement noise. In the absence of detector noise, the measurement SNR will be given by speckle statistics, dictated by the AOT detector's etendue (see Discussion section). In any realistic experimental system, detector noise will be added to the fundamental limit of speckle statistics. We have numerically studied the effect of added measurement noise on the second-order cumulant estimation. The results of this study are summarized in Supplementary Figure S4.

Figure S4 shows the results of three numerical experiments where a two-lines object was imaged using the same acoustic PSF and $N_{speckles}$ with various levels of added detector noise. The acoustic focus and the object where taken as identical to those in the experiments of Figure 2 (see Results section). In each simulated experiment, 200 realizations of random speckle illumination patterns were calculated, with a speckle grain size of $\sim 2.25 \mu m$, resulting in $N_{speckles} \approx 4,000$. Various levels of Gaussian noise were added to each of the measurements with average power of $1\% - 10\%$ of the average detected total power. For each experiment (i.e. noise level) the second cumulant was estimated at each spatial position of the ultrasound focus, and the results of the square root of the second cumulant trace were plotted as a function of ultrasound focus position. As can be seen from Figure S4, measurement noise results in an increased background as well as increased estimation error at the second cumulant. When the measurement noise is strong enough it effectively masks the fluctuations signals. In our experimental system, measurement noise was approximately $3\% - 5\%$ of the total detected modulated power.

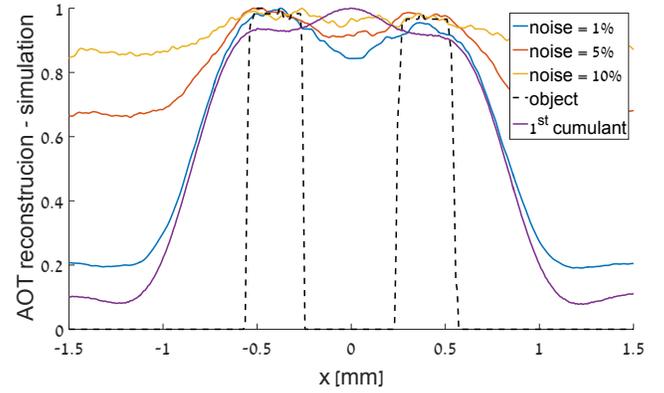

**Fig. S4.** Numerical investigation of the effect of measurement noise on second-order cumulant estimation.


\pdfoutput=1

\end{document}